\documentclass[a4paper,12pt,twoside]{article}
\usepackage{latexsym}
in25mm
\else\oddsidemargin25mm\evensidemargin25mm\marginparwidth25mm\fi
\begin{document}
\def\a{\alpha}
\def\b{\beta}
\def\g{\gamma}
\def\d{\delta}
\def\e{\epsilon}
\def\ve{\varepsilon}
\def\t{\theta}
\def\l{\lambda}
\def\m{\mu}
\def\n{\nu}
\def\pg{\pi}
\def\r{\rho}
\def\s{\sigma}
\def\t{\tau}
\def\c{\chi}
\def\p{\psi}
\def\o{\omega}
\def\G{\Gamma}
\def\D{\Delta}
\def\T{\Theta}
\def\L{\Lambda}
\def\Pg{\Pi}
\def\S{\Sigma}
\def\O{\Omega}
\def\pb{\bar{\psi}}
\def\cb{\bar{\chi}}
\def\lb{\bar{\lambda}}
\def\i{\imath}
\def\eq#1{(\ref{#1})}
\newcommand{\be}{\begin{equation}}
\newcommand{\ee}{\end{equation}}
\newcommand{\ba}{\begin{eqnarray}}
\newcommand{\ea}{\end{eqnarray}}
\newcommand{\ban}{\begin{eqnarray*}}
\newcommand{\ean}{\end{eqnarray*}}
\newcommand{\nn}{\nonumber}
\newcommand{\nin}{\noindent}

\begin{titlepage}
\vskip 0.5cm \hbox{march, 2002} \vskip 2cm
\begin{center}
{\LARGE {$D=6$, $N=2$, $F(4)$-Supergravity\\ with supersymmetric $de\,Sitter$ Background}}\\
\vskip 1.5cm
  {\bf Riccardo D'Auria and Silvia Vaul\`a} \\

\vskip 0.5cm {\small Dipartimento di Fisica, Politecnico di
Torino,\\
 Corso Duca degli Abruzzi 24, I-10129 Torino\\
and Istituto Nazionale di Fisica Nucleare (INFN) - Sezione di
Torino, Italy}\\
\vspace{6pt}
\end{center}
\vskip 3cm
\begin{abstract}
We show that there exists a supersymmetric $de\,Sitter$
background for the\\ $D=6$, $N=2$, $F(4)$ supergravity preserving
the compact $R$--symmetry and gauging with respect to the
conventional $Anti\,de\,Sitter$ version of the theory. We
construct the gauged matter coupled $F(4)$ $de\,Sitter$
supergravity explicitly and show that it contains ghosts in the
vector sector.
\end{abstract}

\end{titlepage}

\section{Introduction}
Among the various supergravity theories with an arbitrary number
$N\leq32$ of supersymmetry charges and in arbitrary dimensions
$D\leq11$, the $D=6$, $N=2$, $F(4)$ supergravity plays a special
role. First of all it is based on an exceptional supergroup not
belonging to the $Osp(n|m)$ or $SU(n|m)$ series, namely the $F(4)$
supergroup, according to the classification given in references
\cite{nahm}, \cite{Kac}. Besides, $F(4)$ appears to be the only
supergroup admitting two real sections whose bosonic (even)
generators span either the algebra of $SO(2,5)\otimes SU(2)$ or
$SO(1,6)\otimes SU(2)$, namely two product groups whose first
factor describes the isometry group of $Anti\,de\,Sitter$ or
$de\,Sitter$ space--time respectively (see for example the
classification given in references \cite{vanpro},
\cite{ledfer}).\\
The above property points at the possibility of constructing a
$D=6$, $N=2$ (sixteen supersymmetries) supergravity theory,
admitting a supersymmetric vacuum whose space--time metric is of
$de\,Sitter$ type, besides the conventional possibility discussed
in references \cite{romans}, \cite{nostro1}, \cite{nostro2} of
having an $anti\,de\,Sitter$ supersymmetric vacuum. Actually, the
latter theory was first constructed by Romans \cite{romans} for
the pure supergravity multiplet, using the Noether approach, and
it was found that the theory admits a supersymmetric
$Anti\,de\,Sitter$ ($AdS_6$) background when the gauge coupling
constant $g$ and the $AdS_6$ radius $R=(2m)^{-1}$ satisfy the
relation $g=3m$. Since Romans' theory is unique by construction,
it is clear that an $F(4)$ supergravity admitting a $de\,Sitter$
background, can only exist if it differs from Romans' theory; in
this respect, we will find that a supersymmetric $de\,Sitter$
background implies necessarily the presence of ghosts.\\
It is a well known fact that in conventional supergravities, quite
generally, $de\,Sitter$ configurations break supersymmetry
completely \cite{rompi}. Actually, several examples have been
given recently of "variant" supergravity theories admitting a
supersymmetric $de\,Sitter$ backgrounds which, besides possessing
non compact $R$--symmetry groups are also plagued with the
presence of ghosts \cite{piaghe}. These "variant" theories, are
string theories in $D=10$ or $M$--theory in $D=11$, which can
have the usual Minkowski signature or a non standard signature of
space--time. They are usually obtained by applying $T$--duality
transformations on timelike circles or Lorentzian tori starting
from conventional theories. Generally the theories thus obtained
have ghosts, but in reference \cite{hull} it was argued that the
corresponding complete string theory could possess some extra
symmetry which could eliminate these spurious degrees of freedom.
In our case we find that the $de\,Sitter$ $F(4)$ supergravity
does indeed show ghosts in the vector sector, however it shows
the peculiarity that the $R$--symmetry group remains compact,
namely it is $SU(2)$ as in the sister $Anti\,de\,Sitter$ theory
of the conventional type \cite{romans}, \cite{nostro1},
\cite{nostro2}. Furthermore, the gauge group of the gauged theory
is exactly the same as for the sister $AdS_6$ theory, differently
from what
happens in the variant theories considered in the literature.\\
In references \cite{nostro1}, \cite{nostro2}, Romans' theory was
reformulated in superspace and the important mechanisms of
generation of a supersymmetric $AdS_6$ configuration and of the
Higgs phenomenon for the generation of a mass term for the
two--form sitting in the gravitation multiplet was explained in
purely algebraic terms, by studying the superspace algebra in its
dual form. Subsequently the theory was coupled to an arbitrary
number $n$ of vector multiplets and the gauging of the gauge
group $SU(2)\otimes\mathcal{G}$, with $\mathcal{G}\subset SO(n)$,
$n$--dimensional compact subgroup, was performed.\newpage The
approach we follow for the construction of the $F(4)$
$de\,Sitter$ supergravity is quite analogous to that used in
references \cite{nostro1}, \cite{nostro2}.\\ We start from the
Maurer--Cartan equations (MCE) equations of the supergroup
$F^t(4)$ containing the proper subgroups $SO(t,7-t)$, with
$t=1,2$. Thus $F^1(4)$ contains the $de\,Sitter$ group and
$F^2(4)$ contains the $Anti\,de\,Sitter$ group in six
dimensions.\\
It is then straightforward to pull--back the MCE of $F^t(4)$ to
the MCE of the coset $F^t(4)/SO(1,5)\otimes SU(2)$ which describes
supersymmetric configurations of vacua whose metric is of
$de\,Sitter$ or $Anti\,de\,Sitter$ type, for $t=1$, $t=2$
respectively. Once this have been obtained, the construction of
the relevant supergravity theory can be performed in the usual
way, defining the supercurvatures out of the vacuum and by using
the Bianchi identities in superspace, as it was extensively
discuss in reference \cite{nostro1} in the $F^2(4)$ case.\\
Since $F^1(4)$ and $F^2(4)$ are two different real sections of
the same complex supergroup $F(4)$, the discussion of the
differences between them has some subtleties related to the
discussion of the different reality properties of the bilinear
fermion currents appearing in the two cases. Even if rather
technical, the discussion of these points is quite essential in
order to understand the relations between the two theories and
the appearance of ghosts in the $de\,Sitter$ case.\\
The plan of the paper is the following:

in section 2 we construct the MCE for the two cosets and their
extension to a Free Differential Algebra (FDA), in order to
include in the game the 2--form $B_{\m\n}$ and the extra $SU(2)$
singlet vector $A_{\m}$ sitting in the gravitational
supermultiplet. A full discussion of the reality properties of
the FDA is also given.

In section 3 we give the explicit form of the $F^1(4)$
supergravity theory admitting a $de\,Sitter$ supersymmetric
vacuum, its supersymmetry transformation laws and the relevant
terms of the Lagrangian, showing in an explicit way the appearance
of ghosts in the vector sector.

In Appendix A, we describe the modifications of the $F^1(4)$
theory when the coupling to $n$ vector multiplets and the gauging
is performed, while Appendix B describes an alternative approach
to the construction of the $F^1(4)$ theory starting the $F^2(4)$
FDA and performing a suitable map.

\section{Construction of $D=6$\\ Supersymmetric $de\,Sitter$ Background}
To obtain a $D=6$ supergravity with a supersymmetric $de Sitter$
background, we begin to consider the complex superalgebra $F(4)$;
it has two real sections, denoted by $F^t(4)$, $t=1,\,2$
containing as even part (bosonic subalgebra)$SO(t,7-t)\times
SU(2)$, that is the $de\,Sitter$ group $SO(1,6)$ and the
$Anti\,de\,Sitter$ group $SO(2,5)$
respectively, times the $R$--symmetry group $SU(2)$.\\
The $F^t(4)$ superalgebra has the following form

\begin{eqnarray}&&[M_{\hat{a}\hat{b}},M_{\hat{c}\hat{d}}]=-\frac{1}{2}(\eta_{\hat{b}\hat{c}}M_{\hat{a}\hat{d}}
+\eta_{\hat{a}\hat{d}}M_{\hat{b}\hat{c}}-\eta_{\hat{b}\hat{d}}M_{\hat{a}\hat{c}}-
\eta_{\hat{a}\hat{c}}M_{\hat{b}\hat{d}})\nonumber\\
&&[T_r,T_s]=-g\epsilon_{rst}T_t\nonumber\\
&&[M_{\hat{a}\hat{b}},Q_{A\a}]=-\frac{1}{4}(\gamma_{\hat{a}\hat{b}})_{\,\,\,\alpha}^{\beta}Q_{A\b}
\nonumber\\
&&[T_r,Q_{A\alpha}]=\frac{i}{2}g\s_{rA}^{B}Q_{B\alpha}\nonumber\\
\label{7al}&&\{Q_{A\alpha},Q_{B\beta}\}=-2i\s^r_{AB}C_{(-)\alpha\beta}T_r
+\epsilon_{AB}(C_{(-)}\g^{\hat{a}\hat{b}})_{\a\b}M_{\hat{a}\hat{b}}
\end{eqnarray}
\noindent Here $M_{\hat{a}\hat{b}}$, $\hat{a},\hat{b}=0\dots6$ are
the $SO(t,7-t)$ generators, preserving the metric
\be\eta_{\hat{a}\hat{b}}=(\underbrace{+\dots+}_{t\,times}\underbrace{-\dots-}_{7-t\,times})\ee
In the following we will keep the notation $"t"$ for any of the
vector timelike indices and we will denote $"s"$ any of the $7-t$
vector spacelike indices of the metric.\\
In equations \eq{7al} $T^r$, $r=1,2,3$ are the $SU(2)$
generators, $\s^{r\ C}_A$ are the usual Pauli matrices and we set
$\s^r_{AB}=\s^r_{BA}\equiv\s^{r\ C}_A\e_{CB}$, where
$\e_{AB}=-\e_{BA}=-\e^{BA}=\e^{AB}$ is the antisymmetric $SU(2)$
tensor which can be used to raise and lower the $SU(2)$ indices
according to the following rules:
\begin{eqnarray}
&&T^{\dots A\dots}=\epsilon^{AB}\ \ T^{\dots\ \ \dots}_{\ \ B}\\
&&T_{\dots A\dots}=T_{\dots\ \ \dots}^{\ \ B}\ \ \epsilon_{BA}
\end{eqnarray}
Note that, since the bosonic subgroup is $SO(t,s)$, $t+s=7$, the
spinor Clifford algebra is seven--dimensional. Indeed $Q_{A\a}$,
$\a=1\dots 8$, $A=1,2$ are the odd generators of the superalgebra
(supersymmetry charges) consisting of seven--dimensional
8-component symplectic--Majorana spinors, and $C_{(-)\a\b}$ is
the symmetric charge conjugation matrix in $D=7$ satisfying
$C_{(-)}^T=C_{(-)}$, $C_{(-)}^2=\bf{1}$. Note that the $D=7$
gamma matrices are $C_{(-)}$--antisymmetric, that is they satisfy
\be
\label{trans}\g_{\widehat{a}}^T=-C_{(-)}^{-1}\g_{\widehat{a}}\,C_{(-)}
\ee We note that the symplectic--Majorana condition holds for
both cases $t=1$, $t=2$ since we have $\r=s-t=5$, $\r=s-t=3$
respectively, so that the symplectic--Majorana condition on the
spinors can be imposed for both signatures \cite{dfl}, \ba
\label{sign16}&&t=1,\,s=6\,\,\,\eta_{\hat{a}\hat{b}}=(+------)\\
\label{sign25}&&t=2,\,s=5\,\,\,\eta_{\hat{a}\hat{b}}=(++-----) \ea
In fact the spinor charges $Q_{A\a}$ satisfy in both cases the
symplectic--Majorana condition \be
\label{sym}\overline{(Q_A)}=\epsilon^{AB}Q_{B}^{\ \ T}C_{(-)} \ee
On the other hand, the definition of the Dirac conjugate spinor
is given by \be\label{bar}\overline{(Q_A)}=(Q_A)^{\dagger}G^{-1}
\ee We have two possible choices for the matrix $G$ in $D=t+s$
dimensions \cite{regge}: the first choice corresponds to the
product of all the timelike gamma matrices, that is:
 \be \label{GI}G_I=\g^0\dots\g^{t-1} \ee
while the second choice corresponds to take the product of all
the spacelike gamma matrices: \be\label{GII}
G_{II}=\g^{D-s}\dots\g^{D-1}\ee Correspondingly the hermitian
conjugate of the gamma matrices is given by: \be
\label{dagger1}\g^{a\dagger}=\eta\, G^{-1}\g^aG \ee where the
phase $\eta$ takes two different values according to the two
choices of $G$, namely \be\label{dagger2}
\eta_I=(-1)^{t-1};\,\,\,\,\,\eta_{II}=(-1)^s\ee

\nin For seven dimensional spinors we choose to use $G_I$ for both
signatures. However the consideration of both conventions will be
relevant when we will reduce the MCE to retrieve the six
dimensional theory.\\
We note that the consistency of the last equation of \eq{7al}
requires an appropriate definition of the hermitian conjugate of
the anticommutator in order that both $T_r$ and
$M_{\hat{a}\hat{b}}$ be antihermitian generators for both
signatures. It is convenient to discuss this point in terms of
the dual formulation of the algebra \eq{7al}. Introducing the
$F^t(4)$ Lie algebra valued 1-forms $\o^{\hat{a}\hat{b}}$, $A^r$,
$\p^{A\a}$ satisfying: \be
\o^{\hat{a}\hat{b}}(M_{\hat{c}\hat{d}})=\d^{\hat{a}\hat{b}}_{\hat{c}\hat{d}},\,\,\,\,\,A^r(T_s)=\d^r_s,\,\,\,\,\,
\p^{A\a}(Q_{B\b})=\d^A_B\d^\a_\b\ee and using the well known
relation \be d\m^A(X_B,X_C)=-\frac{1}{2}\m^A([X_B,X_C])\ee where
$X_B$, $X_C$ are the generators of the super--Lie algebra, and
$\m^A$ are their dual forms, one easily obtains (omitting the
wedge product symbol):
\ba
&&d\o^{\hat{a}\hat{b}}-\o^{\hat{a}}_{\,\ \hat{c}}\o^{\hat{c}\hat{b}}+\frac{1}{2}\pb_A\g^{\hat{a}\hat{b}}\p^A=0\nn\\
&&d\p_A-\frac{1}{4}\o^{\hat{a}\hat{b}}\g_{\hat{a}\hat{b}}\p_A+\frac{i}{2}g\s^r_{AB}A_r\p^B=0\nn\\
\label{mce}&&dA^r-\frac{1}{2}g\e^{rst}A_sA_t-i\s^r_{AB}\pb^A\p^B=0
\ea Note that $\p_A$ satisfies the same properties as $Q_A$,
namely it is a symplectic--Majorana spinor  1-form obeying \be
\label{sympsi}\overline{(\p_A)}\equiv\p^{\dagger}G^{-1}=\epsilon^{AB}\p_{B}^{\
\ T}C_{(-)}\ee When the algebra $F^t(4)$ is written in this form,
one can easily check its closure under $d$-differentiation, which
amounts to the closure of the Jacobi identities in equations
\eq{7al}. One finds that it closes if and only if $g=\frac{3}{2}$.
The verification relies on the following Fierz identity valid for
symplectic--Majorana spinor 1-forms in $D=7$:
\be\label{fierz7}\frac{1}{4}\g_{\hat{a}\hat{b}}\p_A\pb_B\g^{\hat{a}\hat{b}}\p^B+3\p^B\pb_A\p_B=0\ee
We now observe that in arbitrary dimension $D=t+s$, the reality of
the currents $j^{a_1\dots a_n}\equiv\pb_A\g^{a_1\dots a_n}\p_B$
(where $\g^{a_1\dots a_n}$ is a totally antisymmetric product of
gamma matrices) depends on the values of $t$ and $s$ appearing in
the metric $\eta_{ab}$. Indeed we have \cite{regge}
\ba &&(j^{a_1\dots a_n})^*=-\b\c j^{a_1\dots a_n}\nn\\
&&\c_I=(-1)^{\frac{1}{2}(t-n-1)(t-n)}\nn\\
\label{complex}&&\c_{II}=(-1)^{\frac{1}{2}(n-s-1)(n-s)}\ea where
the subscripts $I$ and $II$ refer to the two possible choices for
the matrix $G=\{G_I,\,\,G_{II}\}$ and $\b=\pm 1$ is the arbitrary
phase appearing in the convention one uses for the definition of
the complex conjugate of the product of two spinors, namely
\cite{vanpro}, \cite{regge}  \be (\l\m)^*=\b\l^*\m^*=-\b\m^*\l^*
\ee One can verify that the two currents appearing in equations
\eq{mce} behave under complex conjugation ( using $G=G_I$) as
shown in Table \ref{7curr}: \vskip 0.5cm
\begin{table}[h]
\begin{center}
\begin{tabular}{|r|r|r|l|}
\hline
\,&$\pb_A\p_B$&$\pb_A\g^{\hat{a}}\p_B$&$\pb_A\g^{\hat{a}\hat{b}}\p_B$\\
\hline
$(t,s)=(1,6)$&$\c_I=+1$&$\c_I=+1$&$\c_I=-1$\\
\hline
$(t,s)=(2,5)$&$\c_I=-1$&$\c_I=+1$&$\c_I=+1$\\
\hline
\end{tabular}
\caption{Values of the phase $\c_I$ for $D=7$
currents}\label{7curr}
\end{center}
\end{table}
\vskip 0.5cm Consistency of equations \eq{mce} requires that
$\pb_A\g^{\hat{a}\hat{b}}\p_B$ is real and $\pb_A\p_B$ is pure
imaginary, so that, if we want to have the same algebra for both
signatures, we have to choose $\b=1$ for $(t,s)=(1,6)$ and
$\b=-1$ for $(t,s)=(2,5)$. Alternatively one could use the same
convention for $\b$, say $\b=1$, in which case equations \eq{mce}
would refer to the $(t,s)=(1,6)$ case, while the $(t,s)=(2,5)$
would give analogous equations with an $i$ factor in front of the
two currents. We choose the first alternative and the
corresponding algebra of (anti)--commutators is the one written
in equations \eq{7al} which indeed corresponds
to both choices for $\eta_{\hat{a}\hat{b}}$.\\
The foregoing discussion is propaedeutical to the search of
$de\,\,Sitter$ and $Anti\,\,de\,\,Sitter$ supersymmetric $D=6$
backgrounds. To reach this goal we reduce the seven dimensional
indices $\hat{a},\,\hat{b}=0,1\dots 6$ to six dimensional ones
$a,\,b=0,1\dots 5$ plus the index $6$. Furthermore the reduced
1--forms can be now interpreted as living on the pull-back of
$F^t(4)$ to the coset $F^t(4)/SO(1,5)\otimes SU(2)\supset
SO(t,s)/SO(1,5)$ for both
choices of $t$ and $s$ \eq{sign16}, \eq{sign25}.\\
In order to obtain such backgrounds in $D=6$ we reduce the
indices in such a way that the index "6" is of type $"t"$ for the
signature $(t,s)=(2,5)$, corresponding to an $AdS$ background,
while to obtain a $dS$ background we start from the signature
$(t,s)=(1,6)$ and perform the reduction in such a way that the
index $"6"$ is of type $"s"$. Note that the procedure is
consistent also for the odd part of the algebra, because the
gamma matrices are the same in $D=6$ and $D=7$, and the spinors
in $D=6$ with Minkowski signature $(t,s)=(1,5)$, are still
symplectic-Majorana since $\r=s-t=4$. Furthermore we can still
use $C_{(-)}$ as charge conjugation matrix, since in six
dimensions we have two possible charge conjugation matrices
$C_{(+)}$, $C_{(-)}$ and we can choose $C_{(-)}$ coincident with
the $C_{(-)}$ defined for the $D=7$ spinor algebra. However,
there is a difference between the two cases: reducing on a $"s"$
type direction ($dS_6$ case) the number of timelike gamma
matrices is unchanged, so we can keep the definition of the
barred spinor in $D=6$ as
\be\label{bar1}\pb_A=\p_A^\dagger\g^0\equiv\p_A^{\dagger}G_I^{-1}\ee
On the other hand, if we reduce on a $"t"$ direction ($AdS_6$
case), the seven dimensional definition
$\pb_A=\p_A^{\dagger}(-\g^0\g^6)$ contains the $\g^6$ matrix that
is no more associated to a space-time direction in $D=6$. This is
a crucial point, and the consequences can be understood
considering that in the two cases we have: \ba
&&AdS:\,\,\,\,\eta_{66}=1\Longrightarrow\g_6=\g^6\equiv\g_6^{(+)},\,\,\,\,(\g_6^{(+)})^2=1,\,\,\,\,(\g_6^{(+)})^{\dagger}=\g_6^{(+)}\nn\\
\label{prop+-}&&dS:\,\,\,\,\eta_{66}=-1\Longrightarrow\g_6=-\g^6\equiv\g_6^{(-)},\,\,\,\,(\g_6^{(-)})^2=-1,\,\,\,\,(\g_6^{(-)})^{\dagger}=-\g_6^{(-)}
\ea

\nin From a $D=6$ point of view, the matrix $\g_6$ is usually
called $"\g_7"$ and is defined as
$\g_7=\a\g_0\g_1\g_2\g_3\g_4\g_5$ with $|\a|=1$. If we want to
preserve the properties \eq{prop+-} we choose

\be
\label{defg7}\g_7=\g_6^{(-)}=i\g_6^{(+)}=i\g_0\g_1\g_2\g_3\g_4\g_5
\ee

\nin This means that starting from $D=7$ with $(t,s)=(2,5)$, where

\be \pb_A=\p_A^\dagger(-\g^0\g^6_{(+)}) \ee

\nin and reducing on a $"t"$ type direction, the Dirac conjugate
spinor in $D=6$ turns out to be defined as

\be\label{bar2}\pb_A=\p_A^\dagger(-\g^0\g^6_{(+)})=\p_A^{\dagger}(i\g^0\g^7)=\p_A^{\dagger}(-\g_1\g_2\g_3\g_4\g_5)\equiv\p_A^{\dagger}G_{II}^{-1}\ee

\nin that is, we must use convention $II$ to define the Dirac
conjugate
spinor.\\
We can now perform the pull--back to the six dimensional
superspace. Note that since we are doing the pull-back
$F^t(4)\longrightarrow F^t(4)/SO(1,5)$ the spin connection
components $\o^{a6}$ assume the meaning of the vielbein cotangent
frame on the coset. Introducing a (real) rescaling parameter $m$
we set:

\ban
&&\o^{a6}\longrightarrow 2mV^a\\
&&\p_A\longrightarrow\sqrt{2m}\p_A\\
&&A_r\longrightarrow 2mA_r\\
&&g\longrightarrow\frac{1}{2m}g
\ean

\nin so that the M.C.E. \eq{mce} reduce to the following form:

\begin{eqnarray}
\label{blabla1}&&\mathcal{D}V^{a}+\frac{1}{2}\ \
\overline{\psi}_{A}\gamma^{a}\g^{6(\pm)}\psi^A=0 \\
\label{blabla2}&&\mathcal{R}^{ab}+4m^{2}\ \
V^{a}V^{b}\eta_{66}+m\overline{\psi}_{A}\gamma^{ab}\psi^A=0\\
\label{blabla3}&&dA^{r}-\frac{1}{2}\,
g\,\epsilon^{rst}A_{s}A_{t}-i\,\overline{\psi}_{A}\psi_{B}\,\
\sigma^{rAB}=0\\
\label{blabla4}&&D\psi_{A}+m\gamma_{a}\g_6^{(\pm)}\psi_{A}V^{a}=0\end{eqnarray}

\nin where $\mathcal{R}^{ab}$ is the Lorentz curvature 2--form in
$D=6$, namely

\be\mathcal{R}^{ab}=d\o^{ab}-\o^a_{\,\,c}\o^{cb}\ee

\nin and the $SU(2)$ vector field strength is defined as

\be\mathcal{F}^r=dA^{r}-\frac{1}{2}\,
g\,\epsilon^{rst}A_{s}A_{t}\ee

\nin Moreover, we have defined the Lorentz covariant derivatives
in $D=6$ as follows \ba
&&\mathcal{D}V^{a}=dV^a-\o^{ab}V_b\\
&& \mathcal{D}\p_A=d\p_A-\frac{1}{4}\g_{ab}\,\o^{ab}\p_A \ea

\nin and the $SU(2)\times SO(1,5)$ covariant derivative on $\p_A$
as

\be D\p_A=\mathcal{D}\p_A+\frac{i}{2}\s^r_{AB}A_r\p^B\ee

Equations \eq{blabla1}--\eq{blabla4} define a supersymmetric
vacuum configuration of $D=6$, $N=2$ supergravity in terms of the
superforms $V^a$, $\o^{ab}$, $A_r$, $\p_A$.\\
If we now perform $d$-differentiation of equations
\eq{blabla1}--\eq{blabla4}, we find that they of course close,
since they are merely a rewriting of equations \eq{mce}; however
the relevant Fierz identity to be used in this case, namely in
$D=6$ is the following

\be\label{fierz6}\frac{m}{4}\g_{ab}\p_A\pb_B\g^{ab}\p^B+g\p^B\pb_A\p_B+\frac{m}{2}\g_6\g_a\p_A\pb_B\g^6\g^a\p^B=0\ee

\nin which only holds when $g=3m$. Indeed in this case equation
\eq{fierz6} reduces to equation \eq{fierz7} written in six
dimensional formalism.\\
Note that the dependence of this equations from the signature
$(t,s)$ is hidden in the different values of $\eta_{66}$ and
properties of $\g_6^{(\pm)}$ as given in equations \eq{prop+-}.\\
In particular, we see that, restricting the forms to space-time,
so that $\p_{A\m}=0$, equation \eq{blabla2} gives \be
\mathcal{R}^{ab}_{cd}=-4m^2\eta_{66}\d^{ab}_{cd}\longrightarrow\mathcal{R}_{\m\n}=20\eta_{66}m^2g_{\m\n}
\ee
so that, depending on the signature $(t,s)=(2,5)$ or
$(t,s)=(1,6)$, we get a supersymmetric $Anti\,de\,Sitter$ or
$de\,Sitter$ background with cosmological constant $\pm 20m^2$
respectively. As we stressed in the introduction, the fact that
we can obtain a supersymmetric $de\,Sitter$ configuration is a
peculiar property of the six dimensional theory based on the
$F(4)$ supergroup which admits two real sections, one with
bosonic subgroup $SO(2,5)\times SU(2)$ and the other with bosonic
subgroup $SO(1,6)\times SU(2)$.\\
On the other hand, the supersymmetric configuration described
before, is not complete, since the $D=6$ $F(4)$ supergravity
theory contains, besides the previously treated degrees of
freedom, also an additional vector $A_{\m}$ and a 2--form
$B_{\m\n}$. The way to introduce these extra fields in the
configuration is well known \cite{nostro1}, \cite{nostro2} and
consists in enlarging the MCE \eq{blabla1}--\eq{blabla4} into a
Free Differential Algebra (FDA). Indeed let us add the following
2--form and 3--form equations to the MCE
\eq{blabla1}--\eq{blabla4}:

\ba\label{blabla5}&&dA-mB+\a\,\overline{\psi}_{A}\gamma_{6}^{(\pm)}\psi^A=0\\
&&\label{blabla6}dB+\b\,\overline{\psi}_{A}\gamma_{a}\psi^AV^{a}=0\ea

\nin where $\a$ and $\b$ are suitable constants. Performing
$d$--differentiation of equations \eq{blabla5}--\eq{blabla6},
using the Fierz identity

\be\overline{\psi}_{A}\gamma_{a}\psi^A\overline{\psi}_{B}\gamma^{6}\g^a\psi^B=0\ee

\nin the MCE \eq{mce} and the fact that and $\g_{ab}\g_6$ is
$C_{(-)}$ symmetric, one easily finds

\be\b=2\a\eta_{66}\ee From Table \ref{7curr} we see that in
signature $(t,s)=(2,5)$ the vector current has the same reality
as the tensor current, therefore it is real; vice versa in
signature $(t,s)=(1,6)$ they have opposite reality, that is the
vector current is pure imaginary. Hence we choose  $\b=2$ and
$\b=2i$ for $(t,s)=(2,5)$ or $(t,s)=(1,6)$ respectively, to
respect the reality of the fields $A$ and $B$.\\
At this point we have all the necessary ingredients to construct
the six dimensional supergravity theories which will admit as
supersymmetric background either a six dimensional $Anti\,de\,
Sitter$ ($(t,s)=(2,5)$) or a $de\,Sitter$ configuration
($(t,s)=(1,6)$). Hereafter we will refer to these two theories as
$AdS_6$ or $dS_6$ supergravity respectively.\\
Actually $AdS_6$, $N=2$, $F(4)$ supergravity which was constructed
by Romans \cite{romans}, by means of the Noether approach, using
the supergravity multiplet only was reformulated in terms of a FDA
in superspace in references \cite{nostro1},\cite{nostro2}.
Furthermore its coupling to matter multiplets and the gauging of
a compact group was given. However, the starting FDA used in
\cite{nostro1},\cite{nostro2} was defined with a different
representation for the gamma matrices. We can map our FDA
\eq{blabla1}--\eq{blabla4}, \eq{blabla5}--\eq{blabla6} into that
of \cite{nostro1},\cite{nostro2} by the following redefinition
which preserves the square of the gamma matrices (and hence the
metric $(t,s)=(2,5)$):
\begin{equation}
\label{ridef1}\gamma^a\longrightarrow
-i\gamma^a\,\gamma^{6(+)}\equiv \gamma^7\,\gamma^a
\end{equation}
\nin With the previous redefinition, the Dirac conjugate spinor in
six dimensions, which was previously defined using convention
$II$ (see equation \eq{bar2}) must be now defined using convention
$I$. In fact, for seven dimensional spinors with $(t,s)=(2,5)$ we
had $\p^{\dagger}G_I^{-1}\equiv\p^{\dagger}(-\g^0\g^{6(+)})$. If
we now go to six dimensions, the spinors do not reduce, and
performing the redefinition \eq{ridef1} we obtain
\be\label{paciugo1}\p^{\dagger}(-\g^0\g^{6(+)})\longrightarrow\p^{\dagger}(-\g_7\g^0\g^{6(+)})=
i\p^{\dagger}\g^0\equiv i\p^{\dagger}G_I^{-1}\ee
and the $AdS_6$
FDA takes the following form:
\begin{eqnarray}
\label{dv+}&&\mathcal{D}V^{a}-\frac{i}{2}\,\overline{\psi}_{A}\gamma^{a}\psi^A=0 \\
\label{r+}&&\mathcal{R}^{ab}+4m^{2}\,V^{a}V^{b}+m\overline{\psi}_{A}\gamma^{ab}\psi^A=0\\
\label{dar+}&&dA^{r}+\frac{1}{2}\,g\,\epsilon^{rst}A_{s}A_{t}-i\,\overline{\psi}_{A}\psi_{B}\,\sigma^{rAB}=0\\
\label{ro+}&&D\psi_{A}-im\gamma_{a}\psi_{A}V^{a}=0\\
\label{da+}&&dA-mB-i\,\overline{\psi}_{A}\gamma_{7}\psi^A=0\\
\label{db+}&&dB+2\,\overline{\psi}_{A}\gamma_{7}\gamma_{a}\psi^AV^{a}=0
\ea
The corresponding supergravity has been thoroughly discussed
in
\cite{nostro1},\cite{nostro2} and we do not dwell on it anymore.\\
Our interest instead is to construct the $de\,Sitter$
supergravity and in order to profit of the results obtained in
the $AdS_6$ case, it is convenient to redefine also for $dS_6$
supergravity the gamma matrix representation as follows:
\be\label{ridef2}\gamma^a\longrightarrow
-\gamma^a\,\gamma^{6(-)}\equiv \gamma^7\,\gamma^a\ee With the
previous redefinition, the Dirac conjugate spinor in six
dimensions must be defined using convention $II$, that is with
opposite convention with respect to equation \eq{bar1}. Indeed,
for seven dimensional spinors with $(t,s)=(1,6)$ we have
$\p^{\dagger}G_I^{-1}\equiv\p^{\dagger}\g^0$. Proceeding as
before and using now \eq{ridef2} we obtain

\be\label{paciugo2}\p^{\dagger}\g^0\longrightarrow\p^{\dagger}\g_7\g^0=\p^{\dagger}(i\g^0\g^1\g^2\g^3\g^4\g^5)\g^0
\equiv i\p^{\dagger}G_{II}^{-1}\ee

The resulting FDA for $dS_6$ supergravity is therefore

\begin{eqnarray}
\label{dv-}&&\mathcal{D}V^{a}+\frac{1}{2}\,\overline{\psi}_{A}\gamma^{a}\psi^A=0 \\
\label{r-}&&\mathcal{R}^{ab}-4m^{2}\,V^{a}V^{b}+m\overline{\psi}_{A}\gamma^{ab}\psi^A=0\\
\label{dar-}&&dA^{r}+\frac{1}{2}\,g\,\epsilon^{rst}A_{s}A_{t}-i\,\overline{\psi}_{A}\psi_{B}\,\sigma^{rAB}=0\\
\label{ro-}&&D\psi_{A}-m\gamma_{a}\psi_{A}V^{a}=0\\
\label{da-}&&dA-mB-i\,\overline{\psi}_{A}\gamma_{7}\psi^A=0\\
\label{db-}&&dB-2i\,\overline{\psi}_{A}\gamma_{7}\gamma_{a}\psi^AV^{a}=0
\end{eqnarray}

As a check of our computation we can verify that all the terms
involving bilinear currents in equations \eq{dv+}--\eq{db+},
\eq{dv-}--\eq{db-} are real, consistently with the fact that the
bosonic fields must be real. Indeed, taking into account that the
Dirac conjugate of the spinors in both cases have an extra $i$
factor with respect to the usual definition \eq{paciugo1},
\eq{paciugo2}, that the $\g_7$ matrix has an explicit $i$ factor
\eq{defg7} and using furthermore the relation \be\g_7\g_{a_1\dots
a_p}=\frac{i}{(D-p)!}\e_{a_1\dots a_p\,b_{p+1}\dots
b_D}\g^{b_{p+1}\dots b_D}\ee one obtains the phases shown in Table
\ref{6curr} for the relevant six dimensional currents

\vskip 0.5cm
\begin{table}[h]
\begin{center}
\begin{tabular}{|r|r|r|r|r|r|l|}
\hline
\,&$\pb_a\p_B$&$\pb_a\g_a\p_B$&$\pb_a\g_{ab}\p_B$&$\pb_a\g_7\g_a\p_B$&$\pb_a\g_7\p_B$\\
\hline
I&$\c_I=-1$&$\c_I=-1$&$\c_I=+1$&$\c_I=+1$&$\c_I=-1$\\
\hline
II&$\c_{II}=+1$&$\c_{II}=-1$&$\c_{II}=-1$&$\c_{II}=+1$&$\c_{II}=+1$\\
\hline
\end{tabular}
\caption{Values of the phases $\c_I$, $\c_{II}$ for $D=6$
currents}\label{6curr}
\end{center}
\end{table}
\vskip 0.5cm

We can see that if we set, consistently with the choice in $D=7$,
$\b=1$ for the $dS_6$ case and $\b=-1$ for the $AdS_6$ one, all
the terms containing the bilinear currents in equations
(\ref{dv+})--(\ref{db+}), (\ref{dv-})--(\ref{db-}) are real,
implying the reality of the corresponding bosonic terms.\\
A further check is given by the comparison of equations \eq{ro+}
and \eq{ro-}: indeed these two equations can be interpreted as
the vanishing of the $AdS_6$ and $dS_6$ covariant derivatives.
They differ by a factor $-i$ in the terms proportional to $m$ and
one can wonder whether they are both consistent with the
symplectic-Majorana condition \eq{sym}. The answer is yes; indeed,
let us consider a covariant derivative $\nabla$ defined with a
parameter $q$, where $q=im$ for $AdS_6$ and $q=m$ for $dS_6$:

\be \nabla\p_A=D\p_A-q\g_a\p_AV^a \ee

\nin and let us  apply the symplectic--Majorana condition
\eq{sympsi} to the spinor $\nabla\p_A$; using equations
\eq{trans} and \eq{dagger1}, we have:

\ba
&&\overline{\nabla\p_A}=D\p_A^{\dagger}G^{-1}-q^*\p_A^{\dagger}G^{-1}G\g_a^{\dagger}G^{-1}V^a
=D\overline{\p}_A-q^*\eta\,\overline{\p}_A\g_aV^a\nn\\
&&\e^{AB}\nabla\p_B^TC_{(-)}=\e^{AB}D\p_B^TC_{(-)}-q\e^{AB}\p_B^TC_{(-)}C_{(-)}^{-1}\g_a^TC_{(-)}V^a=D\pb_A+q\pb_A\g_aV^a\nn
\ea
Equating the two r.h.s, and using equation \eq{dagger2} we see
that the only two consistent choices are: $q=im$ using convention
$I$, that is in the $AdS_6$ case, and $q=m$ using convention
$II$, that is in the $dS_6$ case.

\section{$D=6$ $de\,Sitter$ supergravity: transformation rules and
Lagrangian}
In this section we give the form of the supersymmetry
transformation laws and the Lagrangian of $dS_6$ supergravity.\\
Since the underlying algebra is quite analogous to the $AdS_6$
algebra, it is simple to transform the result of the $AdS_6$
theory into the present case. Indeed it is easy to verify (see
Appendix B) that formally we can map the $AdS_6$ FDA
\eq{dv+}--\eq{db+} into the $dS_6$ FDA \eq{dv-}--\eq{db-} by the
following substitutions: \ba
m&\longrightarrow& -im\nn\\
g&\longrightarrow& -ig\nn\\
A^r&\longrightarrow& iA^r\nn\\
A&\longrightarrow& iA\nn\\
B&\longrightarrow& -B\nn\\
\p_A&\longrightarrow& \p_A\nn\\ \label{map}\pb_A&\longrightarrow&
i\pb_A\ea

\nin Out of the vacuum we must also perform the following
substitutions to get the correct form of the superspace
curvatures and hence of the supersymmetry transformation laws:
\ba
\s&\longrightarrow&\s\nn\\
\c_A&\longrightarrow& -i\c_A\nn\\
\overline{\c}_A&\longrightarrow& i\overline{\c}_A\nn\\
\l^I_A&\longrightarrow& -i\l^I_A\nn\\
\overline{\l}^I_A&\longrightarrow& i\overline{\l}^I_A\nn\\
\label{map2}A^I&\longrightarrow& iA^I\nn\\
\ea where $\c_A$ is a spin $\frac{1}{2}$ field and $\s$ a scalar
field belonging to the graviton multiplet, $A^I$, $\l^I_A$,
$P^I_{\a}$ ($\a=1,2,3,4$, $I=1,\dots n$) are the vectors, spin
$\frac{1}{2}$ and scalars of the $n$ vector multiplets (the
scalars are described by the vielbein of the coset
$SO(4,n)/SO(4)\otimes SO(n)$). These substitutions, besides
reproducing the $dS_6$ FDA \eq{dv-}--\eq{db-}, and the
supersymmetry transformation laws out of the vacuum given below,
leave unchanged the gauge invariance \be B\longrightarrow
B+d\l,\,\,\,A\longrightarrow A+m\l \ee
the $R$--symmetry and the Fierz identity \eq{fierz6}.\\
Using the map \eq{map} we can recover the $dS_6$ supersymmetry
transformation laws and Lagrangian by applying it to the
corresponding transformation laws and Lagrangian of $AdS_6$
theory. We have checked that the resulting theory coincides with
the one obtained performing explicit calculation of the Bianchi
identities and of the Lagrangian in the geometric approach. Note
that, equation \eq{map}, \eq{map2} tells us that the kinetic terms
of the vectors undergo a changes of sign with respect to the
$AdS_6$ theory, implying that they behave as ghosts and also the
cosmological constant undergoes a change of sign. This was to be
expected since if one applies the Noether method to construct the
$D=6$ $F(4)$ theory as in reference \cite{romans} the resulting
theory, with positive definite vector kinetic
terms, is unique and of $AdS_6$ type.\\
We note however that, differently from what happens to variant
supergravities discussed in the literature \cite{hullu} the
$R$--symmetry group of our $dS_6$ supergravity remains $SU(2)$,
that is it remains compact compared to the conventional $AdS_6$
supergravity. This is consistent with the fact that all the three
vectors $A^r$
in the adjoint of $SU(2)$ have become ghosts in the $dS_6$ theory.\\
For the sake of clarity we begin to give the supersymmetry
transformation laws for the theory without matter coupling, where
all the important changes with respect to the $AdS_6$ theory
already take place. We have

{\setlength\arraycolsep{1pt}\begin{eqnarray}\label{dva}&\delta
V^{a}_{\mu}&=\overline{\psi}_{A\mu}\gamma^{a}\varepsilon^A\\
&\delta B_{\mu\nu}&=-4i\,\
e^{-2\sigma}\overline{\chi}_{A}\gamma_{7}\gamma_{\mu\nu}\varepsilon^A+4ie^{-2\sigma}\overline{\varepsilon}_A\gamma_7\gamma_{[\mu}\psi_{\nu]}^A\\
 &\delta A_{\mu}&=- 2i\,\ e^{\sigma}
 \overline{\chi}_{A}\gamma_{7}\gamma_{\mu}\varepsilon^A
 +2ie^{\sigma}\overline{\varepsilon}_A\gamma^7\psi^A_{\mu}\\
 &\delta A^{r}_{\mu}&=-2ie^{\sigma}
 \overline{\chi}^{A}\gamma_{\mu}\varepsilon^{B}
\sigma^{r}_{AB}+2ie^{\sigma}\sigma^{rAB}\overline{\varepsilon}_A\psi_{B\mu}\\
&\delta\psi_{A\mu}&=D_{\mu}\varepsilon_{A}+\frac{i}{16}
e^{-\sigma}[\epsilon_{AB}F_{\nu\lambda}\gamma_{7}- \sigma_{rAB}
F^{r}_{\nu\lambda}](\gamma_{\mu}^{\nu\lambda}-6\delta_{\mu}^{\nu}\gamma^{\lambda})\,\
\varepsilon^{B}+\nonumber\\
&&-\frac{i}{32} e^{2\sigma}H_{\nu\lambda\sigma}\,\
\gamma_{7}(\gamma_{\mu}^{\,\ \nu\lambda\sigma}-3\delta_{\mu}^{\nu}
\gamma^{\lambda\sigma}) \varepsilon_{A}-\frac{1}{4}g\,\ e^{\sigma}
\gamma_{\mu}\varepsilon_{A} -\frac{1}{4}
me^{-3\sigma}\gamma_{\mu}\varepsilon_A\nonumber\\
&\delta\chi_{A}&= -\frac{1}{2}
\gamma^{\mu}\partial_{\mu}\sigma\varepsilon_{A}- \frac{i}{16}
e^{\sigma}[\epsilon_{AB} F_{\mu\nu}\gamma_{7}+\sigma_{rAB}
F^{r}_{\mu\nu}]\gamma^{\mu\nu}\varepsilon^{B}\nonumber\\
&&-\frac{i}{32} e^{2\sigma}
H_{\mu\nu\lambda}\gamma_{7}\gamma^{\mu\nu\lambda}\varepsilon_{A}-\frac{1}{4}g
e^{\sigma}\varepsilon_{A}+\frac{3}{4} me^{-3\sigma}\varepsilon_A\\
\label{ds}&\delta\sigma&=\overline{\chi}_{A}\varepsilon^A
\end{eqnarray}
From the transformation laws  \eq{dva}--\eq{ds}, it is easy to see
that one can obtain a $de Sitter$ supersymmetric background
choosing $g=3m$. (Recall that in the vacuum, besides putting to
zero the field-strengths, we also set $\s =0$). In this way we
obtain:
\begin{eqnarray}
&&\delta\chi_A\equiv -\frac{1}{4} \left(g-3m\right)\varepsilon_A= 0\\
&&\delta\psi_{A\mu}\equiv D_\mu \varepsilon_A -(\frac{1}{4}m+\frac{1}{4}g)\gamma_\mu \varepsilon_A=D_\mu \varepsilon_A -m\gamma_\mu \varepsilon_A=\nabla^{dS}_{\mu}\epsilon_A\\
\label{curv}&&R^{ab}\equiv -\frac{1}{2} R^{ab}_{cd}V^cV^d=4m^2 V^aV^b\rightarrow R_{\mu\nu}=-20m^2g_{\mu\nu}\\
&&\left(\mathcal{F}^r_{\mu\nu}=\mathcal{F}_{\mu\nu}-mB_{\mu\nu}=H_{\mu\nu\rho}=\chi_A=\psi_{A\mu}=\sigma=0\right)
\end{eqnarray}
 which corresponds to a $dS$ configuration with $dS$ radius
$R^2_{dS}=(4m^2)^{-1}$

\ba (detV)^{-1}\mathcal{L}&=&
-\frac{1}{4}\mathcal{R}+\frac{1}{8}e^{-2\sigma}[\widehat{\mathcal{F}}_{\mu\nu}\widehat{\mathcal{F}}^{\mu\nu}
+\mathcal{F}^{r}_{\mu\nu}\mathcal{F}^{\mu\nu}_r]
+\frac{3}{64}e^{4\s}H_{\mu\nu\rho}H^{\mu\nu\rho}+\nonumber\\
&&-\frac{1}{2}\overline{\psi}_{A\mu}\gamma^{\mu\nu\rho}
\nabla_{\nu}\psi^{A}_{\rho}+
2\overline{\chi}_A\gamma^{\mu}\nabla_{\mu}\chi^A
+\partial^{\mu}\sigma\partial_{\mu}\sigma+\nn\\
&&-2(-\frac{1}{4}ge^{\s}-\frac{1}{4}me^{-3\s})\overline{\psi}_{\mu
A}\gamma^{\mu\nu}\psi_{\nu}^A
+4(-\frac{1}{4}ge^{\s}+\frac{3}{4}me^{-3\s})\overline{\psi}_{\mu
A}\gamma^{\mu}\chi^A+\nn\\
&&-\mathcal{W}^{ds}(\sigma;\,g,m) \ea

where \ban&&\mathcal{W}^{ds}=g^2e^{2\s}+4mge^{-2\s}-m^2e^{-6\s}\\
&&\widehat{\mathcal{F}}_{\mu\nu}\equiv\mathcal{F}_{\mu\nu}-mB_{\mu\nu}
\ean

We note the "wrong" sign in the vector kinetic term is in
agreement with our previous discussion. If we count the number of
bosonic ghosts and non ghosts degrees of freedom, we find that
they are $16+16$ in agreement with the considerations of
reference \cite{sergio}.\\
We further observe that $\mathcal{W}^{dS}=-\mathcal{W}^{AdS}$ so
that $\mathcal{W}^{dS}$ has the same critical points of
$\mathcal{W}^{AdS}$ for $g=3m$. However, while in the $AdS_6$
case the critical point is a maximum, the dilaton mass is
negative (but satisfies the Breitenlhoner--Friedman bound), in
our case instead we have a minimum of the potential and positive
mass for the dilaton ($m_{\s}^2=24m^2$ or $m_{\s}^2=6$ in
$de\,Sitter$ radius units). We also note that since the vacuum is
supersymmetric, the corresponding extremum is stable.\\ The above
considerations can be straightforwardly extended to the
$de\,Sitter$ supergravity coupled to an arbitrary number of
vector multiplets; one expects that in this case also the vectors
of the vector multiplets become ghosts and again the scalar
potential has a reversed sign with respect to the $AdS_6$ case.
This is in fact what happens and the complete supersymmetry
transformation rules and the Lagrangian are given in the Appendix
A.

\section{Acknowledgements}
We thank Sergio Ferrara for useful discussions and for a critical
reading of the manuscript. One of us (S.V.) acknowledeges J. de Boer for useful discussions.\\
Work supported in part by the European Community's Human Potential Programme under contract HPRN-CT-200-00131 Quantum Spacetime, in which R. D'Auria and S. Vaul\'a are associated to Torino University. 

\appendix
\section*{Appendix A: The gauged matter coupled theory }
\setcounter{equation}{0} \label{appendiceA}
\addtocounter{section}{1}

In this appendix we will briefly discuss the structure of the gauged matter coupled theory; for a more detailed explanation on the complete procedure see \cite{nostro1}, \cite{nostro2}.\\
The only kind of supersymmetric matter in
$D=6$, $N=2$ is given by the vector multiplets, which contain the following fields:
\begin{equation}
(A_{\mu},\,\ \lambda_A,\,\ \phi^{\alpha})^I
\end{equation}
where $\alpha=0,1,2,3$ and the index $I$ labels an arbitrary
number $n$ of such multiplets.\\
With $F^I_{\m\n}$ we will indicate the field strengths associated to the vector fields $A_{\m}^I$; since
the duality group is $G=SO(4,n)\times O(1,1)$, they can be arranged with the graviton multiplet field
strength $F_{\m\n}$, $F^r_{\m\n}$ into an unique $SO(4,n)$ vector $F^{\L}_{\m\n}$ with $\L=0,1,2,3\dots3+n$,
transforming in the fundamental representation of $SO(4,n)$.\\
The $4n$
scalars parametrize the coset manifold $SO(4,n)/SO(4)\times
SO(n)$; they appear in the supersymmetry transformation rules by means of the coset representative $L^{\Lambda}_{\
\ \Sigma}$ of the matter coset manifold, where
$\Lambda,\Sigma,\dots=0, \dots, 3+n$; decomposing the $SO(4,n)$
indices with respect to $H=SO(4)\times SO(n)$ we have:
\begin{equation}
L^{\Lambda}_{\ \ \Sigma}=(L^{\Lambda}_{\ \ \alpha},L^{\Lambda}_{\
\ I})
\end{equation}
\noindent where $\alpha=0,1,2,3$ and $I=4,\dots ,3+n$.
Furthermore, since we are going to gauge $SU(2)\otimes\mathcal{G}\subset SO(4)\times SO(n)$, where $\mathcal{G}$ is an $n$-dimensional subgroup of $SO(n)$ and $SU(2)$ is the diagonal
subgroup of $SO(4)$ as in pure supergravity, we will also
decompose $L^{\Lambda}_{\ \ \alpha}$ as
\begin{equation}
L^{\Lambda}_{\ \ \alpha}=(L^{\Lambda}_{\ \ 0}, L^{\Lambda}_{\ \
r}), \quad \mbox{with  } r=1,2,3.
\end{equation}
The cotangent frame on the scalar coset manifold is given by the vielbein $\widehat{P}^I_{\
0}$, $\widehat{P}^I_{\,\ r}$. They are the $(I,\a)$ components of the gauged left invariant 1-form $L^{-1}\nabla L$
\begin{eqnarray} \widehat{P}^I_{\
0}&=&\left(L^{-1}\right)^I_{\ \L} \nabla L^\L_{\,\ 0}\nonumber\\
\widehat{P}^I_{\,\ r}&=&\left(L^{-1}\right)^I_{\,\ \L} \nabla
L^\L_{\,\ r}.
\end{eqnarray}
\nin where the covariant derivative is defined as
\begin{equation}
\label{nabla}\nabla L^{\Lambda}_{\ \ \Sigma}=d L^{\Lambda}_{\ \
\Sigma}-f_{\Gamma\ \ \Pi}^{\,\ \Lambda} A^{\Gamma} L^{\Pi}_{\ \
\Sigma}
\end{equation}
\noindent where $f^{\Lambda}_{\ \ \Pi\Gamma}$ are the structure
constants of $SU(2)\otimes\mathcal{G}$\\
For simplicity of notation, it is useful to introduce the "dressed" non abelian vector field
strengths:
\begin{eqnarray}
&&\hat{T}_{[AB]\mu\nu}\equiv\epsilon_{AB}L^{-1}_{0\Lambda
}\left(F^{\Lambda}_{\mu\nu} -m B_{\mu\nu} \delta_{\Lambda 0}\right)\\
&&T_{(AB)\mu\nu}\equiv\sigma^r_{AB}L^{-1}_{r\Lambda}F^{\Lambda}_{\mu\nu}\\
&&T_{I\mu\nu}\equiv L^{-1}_{I\Lambda }F^{\Lambda}_{\mu\nu}
\end{eqnarray}
The fermionic supersymmetry transformation rules of the gauged
theory, contain shift terms linear in the parameters $g$, $g'$
and  $m$ (respectively, the gauge coupling constants of  $SU(2)$
and $\mathcal{G}$ and (one half) the inverse of the $dS_6$
 radius); they have been computed for the gauged matter coupled $AdS_6$ $F(4)$ theory in references \cite{nostro1},
  \cite{nostro2}, to which we refer the interest reader for notations and definitions. The fermionic gauge shifts
  were denoted as follows:
\ba &&\d\p_{A\m}=\dots S_{AB}^ {(g,g',m)}\g_{\m}\ve^B\\
&&\d\c_{A}=\dots N_{AB}^ {(g,g',m)}\ve^B\\
&&\d\l^I_A=\dots M_{AB}^ {I\,\,\,\,(g,g',m)}\ve^B \ea We find
that the matrices $N_{AB}^ {(g,g',m)}$ and $M_{AB}^
{I\,\,\,\,(g,g',m)}$, in the present $dS_6$ case, are exactly the
same as before while the gravitino shift acquires a factor $-i$,
so that we define a new matrix $\S^{(g,g',m)}_{AB}$ which is given
in terms of the old one by \be
\S^{(g,g',m)}_{AB}=-iS^{(g,g',m)}_{AB}\ee The resulting
supersymmetry transformation rules for the gauged matter coupled
$F(4)$ $dS_6$ supergravity are therefore:
\begin{eqnarray}
&\delta V^{a}_{\mu}&=\overline{\psi}_{A\mu}\gamma^{a}\varepsilon^A\nonumber\\
&\delta B_{\mu\nu}&=-4ie^{-2\sigma}\overline{\chi}_{A}\gamma_{7}\gamma_{\mu\nu}\varepsilon^A
+i4e^{-2\sigma}\overline{\varepsilon}_A\gamma_7\gamma_{[\mu}\psi_{\nu]}^A\nonumber\\
&\delta A^{\Lambda}_{\mu}&=-2i e^{\sigma}
 \overline{\varepsilon}^{A}\gamma_{7}\gamma_{\mu}\chi^BL^{\Lambda}_0
 \epsilon_{AB}-2ie^{\sigma}\overline{\varepsilon}^{A}\gamma_{\mu}\chi^{B}
 L^{\Lambda r}\sigma_{rAB}+i e^{\sigma}L_{\Lambda
I}\overline{\varepsilon}^{A}\gamma_{\mu}\lambda^{IB}\epsilon_{AB}+\nonumber\\
&&+2ie^{\sigma}L^{\Lambda}_0\overline{\varepsilon}_A\gamma^7\psi_B\epsilon^{AB}+
2ie^{\sigma}L^{\Lambda r}\sigma_{r}^{AB}\overline{\varepsilon}_A\psi_B\nonumber\\
&\delta\psi_{A\mu}&=\mathcal{D}_{\mu}\varepsilon_A+\frac{i}{16}
e^{-\sigma}[\hat{T}_{[AB]\nu\lambda}\gamma_{7}-{T}_{(AB)\nu\lambda}](\gamma_{\mu}^{\,\
\nu\lambda}-6\delta_{\mu}^{\nu}\gamma^{\lambda})
\varepsilon^{B}+\nonumber \\
&&-\frac{i}{32}e^{2\sigma} H_{\nu\lambda\rho}
\gamma_{7}(\gamma_{\mu}^{\,\ \nu\lambda\rho}-3\delta_{\mu}^{\nu}
\gamma^{\lambda\rho})\varepsilon_{A}+\S^{(g,g',m)}_{AB}
\gamma_\mu\varepsilon^B +\nonumber\\
&&+\frac{1}{2}\varepsilon_{A}\overline{\chi}^{C}\psi_{C\mu}+
\frac{1}{2}\gamma_{7}\varepsilon_{A}\overline{\chi}^{C}\gamma^{7}\psi_{C\mu}-
\gamma_{\nu}\varepsilon_{A}\overline{\chi}^{C}\gamma^{\nu}\psi_{C\mu}+\gamma_{7}
\gamma_{\nu}\varepsilon_{A}\overline{\chi}^{C}\gamma^{7}\gamma^{\nu}\psi_{C\mu}+\nonumber\\
&&-\frac{1}{4}\gamma_{\nu\lambda}\varepsilon_{A}\overline{\chi}^{C}\gamma^{\nu\lambda}\psi_{C\mu}-
\frac{1}{4}\gamma_{7}\gamma_{\nu\lambda}\varepsilon_{A}\overline{\chi}^{C}\gamma^{7}\gamma^{\nu\lambda}\psi_{C\mu}
\nonumber\\
&\delta\chi_{A}&=-\frac{1}{2} \gamma^{\mu}\partial_{\mu}\sigma
\varepsilon_{A}-
\frac{i}{16}e^{-\sigma}[\hat{T}_{[AB]\mu\nu}\gamma_{7}+T_{(AB)\mu\nu}]\gamma^{\mu\nu}\varepsilon^{B}-\frac{i}{32}e^{2\sigma}
H_{\mu\nu\lambda}\gamma_{7}\gamma^{\mu\nu\lambda}\varepsilon_{A} + \nonumber\\
&&+ N^{(g,g',m)}_{AB}\epsilon^B \nonumber\\
&\delta\sigma&=\overline{\chi}_{A}\varepsilon^A\nonumber\\
&\delta\lambda^{I}_A&=-\hat{P}^I_{ri}\sigma^{r}_{AB}\partial_{\mu}\phi^{i}\gamma^{\mu}\varepsilon^{B}+
\hat{P}^I_{0i}\epsilon_{AB}\partial_{\mu}\phi^{i}\gamma^{7}\gamma^{\mu}\varepsilon^{B}-
\frac{i}{2}e^{-\sigma}T^{I}_{\mu\nu}\gamma^{\mu\nu}\varepsilon_{A} + M^{I (g,g',m)}_{AB}\varepsilon^B \nonumber \\
&\hat{P}^{I}_{0i}\delta\phi^i&=\frac{1}{2}\overline{\lambda}^{I}_{A}\gamma_{7}\varepsilon^A\nonumber\\
&\hat{P}^{I}_{ri}\delta\phi^i&=\frac{1}{2}\overline{\lambda}^{I}_{A}\varepsilon_{B}\sigma_r^{AB}
\label{newsusy}
\end{eqnarray}

The complete supersymmetric gauged Lagrangian can be obtained
with a standard procedure using equations \eq{newsusy} or by
using the map \eq{map}, \eq{map2} to the corresponding $AdS_6$
Lagrangian given in reference \cite{nostro2}. We limit ourselves to
write down the most interesting terms. Our result is:

\ba \mathcal{L}_{\mbox{\tiny{kin}}}&=&
-\frac{1}{4}\mathcal{R}+\frac{1}{8}e^{-2\sigma}\mathcal{N}_{\Lambda\Sigma}
\widehat{\mathcal{F}}^{\Lambda}_{\mu\nu}\widehat{\mathcal{F}}^{\Sigma\mu\nu}
+\frac{3}{64}e^{4\s}H_{\mu\nu\rho}H^{\mu\nu\rho}+\nonumber\\
&&-\frac{1}{2}\overline{\psi}_{A\mu}\gamma^{\mu\nu\rho}
\nabla_{\nu}\psi^{A}_{\rho}+
2\overline{\chi}_A\gamma^{\mu}\nabla_{\mu}\chi^A-
\frac{1}{8}\overline{\lambda}^I_A\gamma^{\mu}\nabla_{\mu}\lambda^A_I+\nonumber\\
&&+
\partial^{\mu}\sigma\partial_{\mu}\sigma  -\frac{1}{4}
\left(\widehat{P}^{I0}_{ i} \widehat{P}_{I0 j}
+\widehat{P}^{Ir}_{i}\widehat{P}_{Ir , j}\right)
\partial^{\mu}\phi^i\partial_{\mu}\phi^{j}
\,;\ea
\ba
 \mathcal{L}_{\mbox{\tiny{Chern--Simons}}}&=&
 -\frac{1}{64}\epsilon^{\mu\nu\rho\s\l\tau}B_{\mu\nu}
\left(\eta_{\L\Sigma}\hat{\mathcal{F}}^{\L}_{\rho\s}\hat{\mathcal{F}}^{\Sigma}_{\l\tau}
+mB_{\rho\s}\hat{\mathcal{F}}^0_{\l\tau}+\frac{1}{3}m^2B_{\rho\s}B_{\l\tau}\right)
\ea
\ba
 \mathcal{L}_{\mbox{\tiny{gauging}}}&=&
-2\overline{\psi}_{\mu}^A\gamma^{\mu\nu}\overline{\S}_{AB}\psi_{\nu}^B+
+4\overline{\psi}_{\mu}^A\gamma^{\mu}\overline{N}_{AB}\chi^{B}-\frac{1}{4}
\overline{\psi}_{\mu}^A\gamma^{\mu}\overline{M}_{AB}^I\lambda^{B}_I+\nonumber\\
&& -\mathcal{W}^{ds}(\sigma , \, \phi^i;g,g',m) .\label{lgaug}\ea
The fermionic gauge shifts and the scalar potential of $
\mathcal{L}_{\mbox{\tiny{gauging}}}$ appearing in equation
\eq{lgaug} have been given in references \cite{nostro1},
\cite{nostro2}. We note that, as it was to be expected from the
covariance of the vectors under duality group $SO(4,n)\otimes
O(1,1)$, also the matter multiplet vector
fields become ghosts.\\
Moreover, also for the matter coupled theory, the potential
$\mathcal{W}^{ds}$ has the opposite sign with respect to the
$AdS_6$ case, $\mathcal{W}^{ds}=-\mathcal{W}^{Ads}$. The dilaton
and the matter scalar fields are not ghosts, but they all have
positive mass, corresponding to a minimum of the potential. The
linearized equations of motion for the scalar fields are:
\begin{eqnarray}
&&\Box\sigma+24m^2\sigma=0\\
&&\Box q^{I0}+16m^2 q^{I0}=0\\
&&\Box q^{Ir}+24m^2 q^{Ir}=0
\end{eqnarray}
\noindent If we use as mass unity the inverse $dS$ radius, which
in our conventions is $R^{-2}_{dS}=4m^2$ we get:
\begin{eqnarray}
&&m^2_{\sigma}=6\nonumber\\
&&m^2_{q^{I0}}=4\nonumber\\
&&m^2_{q^{Ir}}=6
\end{eqnarray}

\section*{Appendix B: An alternative approach to the\newline construction of $dS_6$ $F(4)$ superalgebra}
\setcounter{equation}{0} \label{appendiceB}
\addtocounter{section}{1} In this appendix we give a detailed
discussion and justification with regard to the derivation of the map \eq{map}, \eq{map2} which have been used in the main text.\\
Let us consider the $F(4)$ $AdS_6$ FDA \eq{dv+}--\eq{db+}; in
order to obtain from it a $dS_6$ supersymmetric FDA, the most
important point is of course to change the sign in front of
$4m^2$ in \eq{r+} and to try to adjust the coefficients in the
various terms of the other equations by imposing the $d$--closure
of the new FDA. On the other hand, in the approach given in the
main body of the paper, we have seen that the $AdS_6$ and the
$dS_6$ background can be obtained reducing respectively on an
$''t''$ type direction or on an $''s''$ direction, which implies a
different definitions of the Dirac conjugate spinor \eq{bar2},
\eq{bar1} for the two theories. Note that this is a general
feature of $D$--dimensional theories with the same background
signature $D=t+s$ that are obtained reducing a $D+1$ supergravity
on a $''t''$ type direction, $D+1=(t+1)+s$, or on a $''s''$
type direction, $D+1=t+(s+1)$.\\
To recover the $dS_6$ theory we can now proceed as follows.
Consider the FDA \eq{blabla1}--\eq{blabla4}, \eq{blabla5},
\eq{blabla6} for the $AdS_6$ case, where the Dirac conjugate
spinors are defined according to convention $I$\footnote{We
recall that because of redefinitions \eq{paciugo1},
\eq{paciugo2}, there has been an interchange between convention
$I$ and $II$}. If we now change convention $I$ into convention
$II$ for the Dirac conjugate spinors (according to the reduction
on a $"s"$ type direction), the algebra is formally the same and
closes as well for $g=3m$, but the fact that some currents have
changed reality means that some of the bosonic fields have become
purely imaginary. Actually, fixing $\b=1$, from Table \eq{6curr}
we see that while all the fields are real for the $AdS_6$ case
(following convention $I$), using convention $II$,
$\mathcal{R}_{ab}$, $\mathcal{F}^r$ and
$\widehat{\mathcal{F}}\equiv dA-mB$ are purely imaginary. Since
the derivation of the supersymmetry transformation rules and the
construction of the Lagrangian are not affected by the presence
of purely imaginary fields, we can take for this new theory
(which has both real and imaginary fields) the same supersymmetry
transformation rules and Lagrangian of $F(4)$ $AdS_6$, but using
Dirac conjugate spinors defined according to convention $II$. At
this point we may rewrite the new theory in terms of real fields
only. Of course, this has to be done preserving the $R$--symmetry
group and the gauge invariance \be\label{gauginv}B\longrightarrow
B+d\l,\,\,\,A\longrightarrow A+m\l \ee If $\mathcal{F}^r$ is pure
imaginary we must define a real field strength
$\widetilde{\mathcal{F}}^r=-i\mathcal{F}^r$; the only way to do
it is to define \ba
&&\widetilde{A^r}=-iA^r\nn\\
\label{pippo1}&&\widetilde{g}=ig
\ea

\nin Furthermore, in the limit $m=0$, corresponding to the theory
where $F(4)$ is contracted to a group containing as a bosonic
subgroup $ISO(1,5)\otimes SO(4)$, to preserve the $SO(4)$
$R$--symmetry group, the field $A$ must be redefined as $A^r$,
that is

\be\label{pippo2}\widetilde{A}=-iA\ee

\nin Since the algebra closes for $g=3m$ we are also forced to
define

\be\label{pippo3}\widetilde{m}=im\ee

\nin Finally, since we want to maintain the gauge invariance
\eq{gauginv}, taking into account \eq{pippo2}, \eq{pippo3}, we
define

\be\label{pippo4}\widetilde{B}=-B\ee

\nin Performing redefinitions \eq{pippo1}--\eq{pippo4} in the Lagrangian we obtain that the kinetic term of the new
vector field strengths $\widetilde{\widehat{\mathcal{F}}}\equiv d\widetilde{A}-\widetilde{m}\widetilde{B}$ and
 $\widetilde{F}^r$ have reversed sign with respect to $\widehat{\mathcal{F}}$ and $\mathcal{F}^r$, that is, they are ghosts,
  according to our previous finding.
In addition we see that we do not have to perform any redefinition
on the spin connection $\mathcal{\o}^{ab}$, since according to
the redefinition \eq{pippo3} on $m$ (see equation \eq{r+})
$\mathcal{\o}^{ab}$ is defined as a real field.\\
We stress that, because of redefinition \eq{pippo3} the
cosmological constant, in terms of $\widetilde{m}^2$ has changed
sign with respect to
the $AdS_6$ case we started from. This means that the new theory describes a $dS_6$ supersymmetric background.\\
To complete the discussion we have also to investigate whether the
spin $\frac{1}{2}$ field $\c_A$ and the dilaton field $\s$, that
appear in the FDA out of the vacuum, do need appropriate
redefinitions too. The superspace differential of the dilaton
field $\s$ is defined in the $AdS_6$ as follows (see references
\cite{nostro1}, \cite{nostro2}): \be\label{sigma}d\s=\partial_a\s
V^a+\cb_A\p^A\ee \nin Following Table \eq{6curr} we see that
using convention $II$ the field $\s$ should be pure imaginary;
obviously this this is not the case. In fact, since the dilaton
$\s$ always appears as $e^{\s}$ in the $AdS_6$ transformation
laws, it would be impossible to impose the symplectic--Majorana
condition \eq{sympsi} on the fermions if we replaced
$e^{\s}\longrightarrow e^{i\s}$. This suggests that the dilatino
$\c_A$ should be redefined, in order to have a real scalar field
$\s$, as follows:

\be\label{pippo5}\widetilde{\c_A}=i\c_A\ee

The same kind of reasoning applied to the matter fields, implies
the following redefinitions: \ba&&\widetilde{A}^I=-iA^I\nn\\
\label{pippo6}&&\widetilde{\l}^I_A=i\l^I_A\ea

\nin Equations \eq{pippo1}--\eq{pippo5} map the $F(4)$ theory with $AdS_6$ background into the $F(4)$ theory
 with $dS_6$ background. If we want to retrieve our previous results, obtained by explicit construction from
 the FDA \eq{dv-}--\eq{db-}, another inessential redefinition is needed: the Dirac conjugate of every
 spinor $\m$
 must be written as:

\be\label{pippo7}\widetilde{\overline{\m}}=-i\overline{\m}\ee

\nin This is just because when we wrote the $F^t(4)$ algebra
\eq{mce}, we wanted it to have the same form for both $F^1(4)$ and
$F^2(4)$. Therefore we had to choose, in order to have real
fields, $\b=-1$ for the $(t,s)=(2,5)$ case, from which the
$AdS_6$ theory comes from.\footnote{If we wanted to use $\b=1$
for both theories and have real fields, we had to modify the
$F^2(4)$ algebra \eq{mce} inserting some $i$ factors in front of
the currents; in that case we wouldn't need redefinition
\eq{pippo7} to obtain the $dS_6$ theory.}

The correctness of redefinition \eq{pippo5} can be checked
considering that the variations under supersymmetry of the
fermions $\d\p_A$, $\d\c_A$ and $\d\l^I_A$ must satisfy the
symplectic--Majorana condition \eq{sympsi} as the fermions do. One
can in fact determine the reality of the coefficient in each term
of the fermionic supersymmetry transformation rules \eq{newsusy},
since it depends only on the number $n$ of gamma matrices that it
contains and on the choice of the convention $I$ or $II$ for
defining the Dirac conjugate spinor.\\
In fact, let us evaluate the transposition and the hermitian
conjugation of the product of $n$ gamma matrices in $D=6$ using
equations \eq{trans}, \eq{dagger1}, \eq{dagger2}: \ba
&&(\g_{a_1}\dots\g_{a_n})^T=(-1)^nC_{(-)}^{-1}\g_{a_n}\dots\g_{a_1}C_{(-)}\nn\\
&&(\g_{a_1}\dots\g_{a_n})^{\dagger}=\a G^{-1}\g_{a_n}\dots\g_{a_1}G\nn\\
\label{rulez}&&\a_{I}=1\,\,\,\,\a_{II}=(-1)^n \ea If we want that
$\d\p_A$, $\d\c_A$ and $\d\l^I_A$ satisfy the
symplectic--Majorana condition \eq{sympsi} the following simple
rule, descending from equations \eq{rulez}, must be obeyed: when
$n$ is even the current must appear in the transformation rule
with a real coefficient in both cases $I$ and $II$; when $n$ is
odd, there must be an imaginary coefficient if we use convention
$I$ and a real one if we use convention $II$.\\
This is exactly what we got for the $AdS_6$ case in reference \cite{nostro1}, \cite{nostro2},
and what we just obtained for the $dS_6$ by explicit construction and using redefinitions \eq{pippo1}--\eq{pippo6}.


\begin{thebibliography}{100}

\bibitem{nahm} W. Nahm, V. Rittenberg and M. Scheunert, J. Math. Phys. {\bf
17} (1976) 1640; W. Nahm, Nucl. Phys. {\bf B135} (1978) 149
\bibitem{Kac}V.G. Kac, Commun. Math. Phys. {\bf 53} (1977) 31;
Adv. Math. {\bf 26} (1977) 8; Jou. Math. Phys. {\bf 21} (1980) 689
\bibitem{vanpro}A. Van Proeyen, Lectures in the spring school in Climanesti, Romania, April 1998, hep-th/9910030
\bibitem{ledfer}S. Ferrara, M.A. Lled\'o, hep-th/0112177
\bibitem{romans} L.J. Romans, Nucl. Phys. {\bf B269} (1986) 691
\bibitem{nostro1} R. D'Auria, S. Ferrara and S. Vaul\' a, JHEP {\bf 0010} (2000) 013, hep-th/0006107
\bibitem{nostro2} L. Andrianopoli, R. D'Auria and S. Vaul\' a, JHEP {\bf 0105} (2001) 065, hep-th/0104155
\bibitem{rompi} M. G\"unaydin, L.J. Romans and N.P. Warner, Phys. Lett. {\bf 154B}, n. 4 (1985) 268;
Nucl. Phys. {\bf B272} (1986) 598.
M. Pernici, K. Pilch and P. van Nieuwenhuizen, Nucl. Phys. {\bf B259} (1985) 460. C.M. Hull, Phys. Rev {\bf D30} (1984) 760; Phys. Lett. {\bf 142B} (1984) 39; Phys. Lett. {\bf 148B} (1984) 297; Class. Quant. Grav {\bf 2} (1985) 343. C.M. Hull and N.P. Warner,  Nucl. Phys. {\bf B253} (1985) 650; Nucl. Phys. {\bf B253} (1985) 675
\bibitem{piaghe}K. Pilch, P. van Nieuwenhuizen and M. Sohnius,
Commun. Math. Phys. {\bf 98} (1985) 105;
C.M. Hull, JHEP {\bf 9807} (1998) 021, hep-th/9806146;
\bibitem{hull} C.M. Hull, JHEP {\bf 9807} (1998) 021, hep-th/9806146
\bibitem{dfl}R. D'Auria, S. Ferrara, M. A. Lled\' o and V. S. Varadarajan, J. Geom. Phys. {\bf 40} (2001) 67, hep-th/0010124
\bibitem{regge} T. Regge, The Group Manifold Approach To Unified Gravity, GIFT Seminar 1984:73
\bibitem{hullu} C.M. Hull and B. Julia, Nucl. Phys. {\bf B534} (1998) 147, hep-th/9803239
\bibitem{sergio} S. Ferrara, hep-th/0101123
\end{thebibliography}
\end{document}